\documentclass[runningheads]{llncs}
\usepackage{booktabs}
\usepackage{cite}
\usepackage{amsmath,amssymb,amsfonts}
\usepackage{algorithmic}
\usepackage{algorithm}
\usepackage{graphicx}
\usepackage{textcomp}
\usepackage{epstopdf}
\usepackage{multirow}
\usepackage{booktabs}
\usepackage[table,xcdraw]{xcolor}
\usepackage{float}
\usepackage{subfigure}
\usepackage{graphicx}
\usepackage{epstopdf}
\usepackage{amsfonts}
\usepackage{amsmath}
\usepackage{multirow}
\usepackage{pdfpages}

\usepackage{marvosym}

\begin{document}
\title{Dynamic Community Detection via Adversarial Temporal Graph Representation Learning}

\author{
Changwei Gong \and
Changhong Jing \and
Yanyan Shen \and
Shuqiang Wang
}
\institute{Shenzhen Institutes of Advanced Technology, Chinese Academy of Sciences, Shenzhen
	518000, China\\
	\email{\{cw.gong,ch.jing,yy.shen,sq.wang\}@siat.ac.cn}}
%
%
\maketitle              
\begin{abstract}
Dynamic community detection has been prospered as a powerful tool for quantifying changes in dynamic brain network connectivity patterns by identifying strongly connected sets of nodes. However, as the network science problems and network data to be processed become gradually more sophisticated, it awaits a better method to efficiently learn low dimensional representation from dynamic network data and reveal its latent function that changes over time in the brain network. In this work, an adversarial temporal graph representation learning (ATGRL) framework is proposed to detect dynamic communities from a small sample of brain network data. It adopts a novel temporal graph attention network as an encoder to capture more efficient spatio-temporal features by attention mechanism in both spatial and temporal dimensions. In addition, the framework employs adversarial training to guide the learning of temporal graph representation and optimize the measurable modularity loss to maximize the modularity of community. Experiments on the real-world brain networks datasets are demonstrated to show the effectiveness of this new method.

\end{abstract}
\section{Introduction}
Neuroscience is emerging into a generation marked by a large amount of complex neural data obtained from large-scale neural systems~\cite{ref_13}. The majority of these extensive data are in the form of data from networks that cover the relationships or interconnections of elements within different types of large-scale neurobiological systems. Significantly, these data often span multiple scales (neurons, circuits, systems, whole brain) or involve different data types in neurobiology (e.g., structural networks expressing anatomical connectivity of nerves, functional networks representing connectivity of distributed brain regions associated with neural activity). The brain network consists of anatomical structures segmenting different brain regions and connecting them by functional networks showing their complex neuronal communication and signaling patterns. Attributed to advancements in current imaging techniques and advanced methods of medical image processing~\cite{ref_24}~\cite{ref_25}~\cite{zuo2021multimodal}~\cite{wang2018automatic}, this sophisticated pattern of neural signals may be studied using functional imaging, in which neuronal activity is associated with a variety of behaviors and cognitive functions as well as brain diseases~\cite{ref_as}~\cite{ref_26}~\cite{ref_27}~\cite{ref_28}~\cite{ref_1}. At the same time, network science is the study of complex network representation through theories and techniques of computer science and mathematics. With the convergence of two significant scientific developments in recent years, new techniques and analytical methods in the network science field are emerging for evaluating real-world biological networks~\cite{ref_14}.

\begin{figure}
\includegraphics[width=\textwidth]{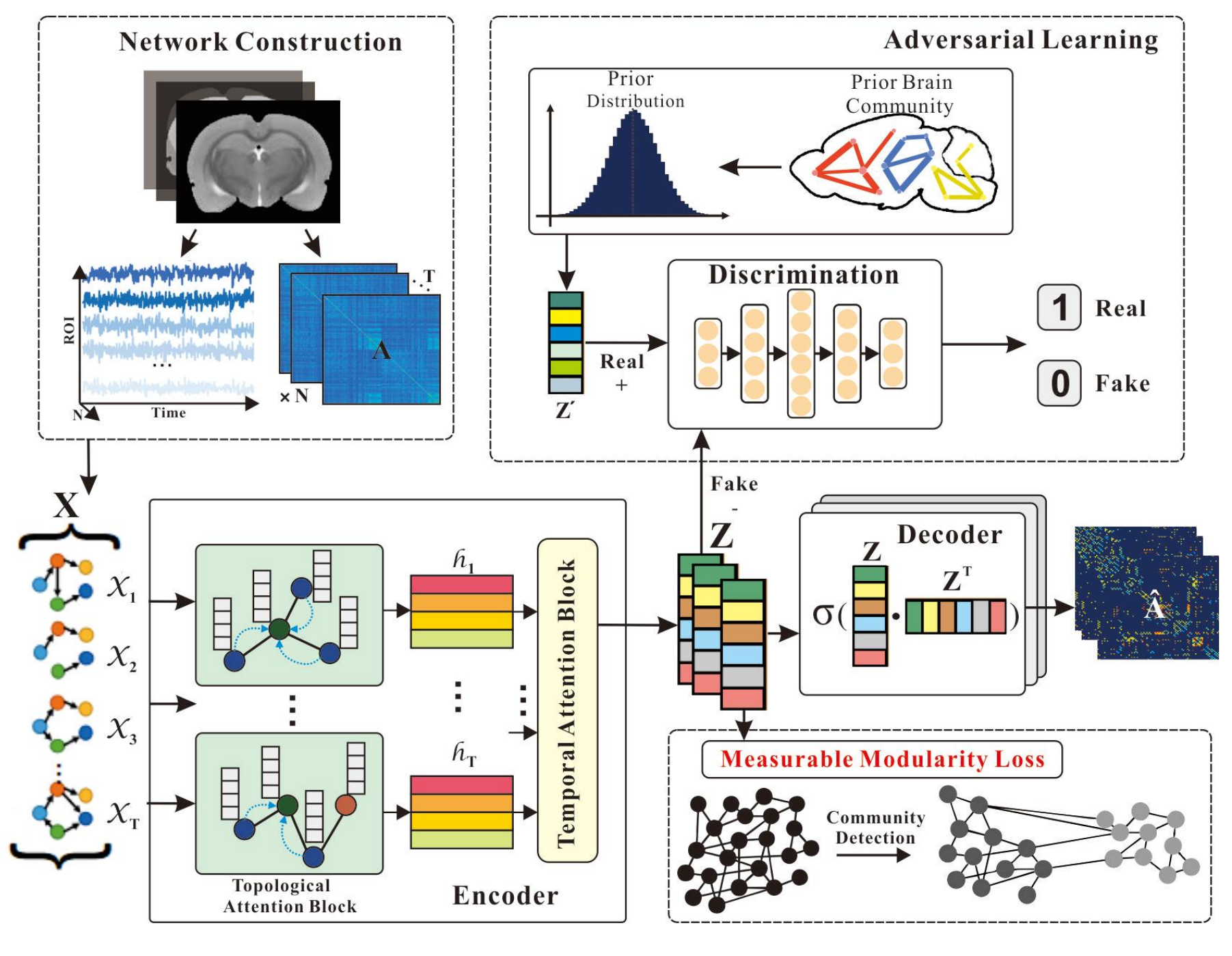}
\caption{Proposed adversarial temporal graph representation learning framework for detecting brain communities.} \label{fig1}
\end{figure}

Subgraphs, network modules, and communities have been extensively studied in the context of network structures, and in particular, community detection~\cite{ref_16}~\cite{ref_31} methods have been widely used in network neuroscience~\cite{ref_15}. Network structure identification or community detection(see schematics in bottom right of Figure.~\ref{fig1}) is the partition of nodes in a network into groups in which nodes in the communities are tightly connected, and nodes in different communities are sparsely connected. The organizational principles and operational functions of complex network systems can be revealed and understood through mining the network structure. Furthermore, the creation of comprehensive network maps of neural circuits and systems has resulted from the development of new techniques for mapping the structure and functional connectivity of the brain. A wide range of graph-theoretic tools can be used to examine and analyze the structure of these brain networks. Therefore, methods for detecting modules or network communities in brain networks are of specialized application, and they reveal tightly connected primary building elements or substructures, which frequently relate to particular functional components.

\subsubsection{Related work.}Data-driven models have gotten much attention for a long time, and when combined with machine learning techniques, it has led to great success in building pattern recognition models within the field of medical image computing~\cite{ref_6}~\cite{ref_7}. The models have the potential to achieve high accuracy at a low computational cost. Deep learning is currently widely perceived as one of the most significant developments in machine learning \cite{wang2018bone, yu2020multi, wang2020ensem}. The method of deep learning is a new approach for dealing with high-dimensional biological data and learning low-dimensional representations of medical image~\cite{ref_4}. The approach based on the generative adversarial methodse~\cite{ref_33} and the graph neural network are good instances. Generative adversarial network(GAN)~\cite{ref_17, wang2007variational,wang2008var}, which can bee seen as variational-inference~\cite{ref_32} based  generative model, is commonly employed in medical image representational analysis~\cite{ref_5}~\cite{ref_8}~\cite{ref_11}. The utilization of GAN for community detection is inspired by the fact that GANs are often supervised in training, and the newly generated data (in principle) has the same distributioun as real data, allowing for robust, complex data analysis~\cite{ref_9}~\cite{ref_10}~\cite{ref_12}. Convolutional neural network (CNN)~\cite{ref_18} approach reduces the dimensions of medical imaging data by pooling and convolution , enabling it to successfully recognize pattern in biomedical task~\cite{ref_2}~\cite{ref_3}. Graph convolutional network (GCN) is developed to extract community features since it derives the CNN capabilities and directly processes on network structured data. However, existing methods to process dynamic network data to obtain temporal graph representations for community detection remain challenging, especially for small sample network datasets.

To end these issues, we developed a novel adversarial temporal graph representation learning (ATGRL) to complete the clustering of brain nodes in dynamic brain networks, detect different communities containing similar brain regions in dynamic brain networks and their evolution, and improve the robustness of the model while handling with small sample network data by employing generative adversarial approaches. The proposed temporal graph attention encoder is efficient to graph representation learning, and more helpful graph embeddings are obtained to complete the clustering to detect more accurate dynamic communities. The detected communities with sound classification effects can be used as biological markers.

\section{Method}
Our ATGRL includes two core parts:1) a temporal graph auto-encoder consists of a temporal graph attention encoder and a decoder, and 2) an adversarial regularizer including a discriminator.The architecture are illustrated in Fig.~\ref{fig1}. In the autoencoder, the encoder ($E$) adopts temporal graph attention networks to transform the time series of brain regions ($\left\{X^t\right\}^T_{t=1}$)  and brain functional connections ($A$) into the embeddings ($\left\{Z^t\right\}^T_{t=1}$) . Moreover, in the adversarial regularizer, a min-max adversarial game is led between the encoder, which regards as the generator ($G$), and the discriminator ($D$) to learn better embeddings. In order to detect communities, the measurable soft modularity loss is employed which optimizes the community assignment matrix $P$. Therefore, the encoder is trained with triple objectives: a classic reconstruction loss of autoencoder, an adversarial training loss from the discriminator and the measurable modularity loss for detecting community.

\subsection{Temporal Graph Autoencoder}
The temporal graph autoencoder aims to embed the dynamic brain network attributes in a low-dimensional latent space. First, we use two different network blocks to construct the encoder: the topological attention block and temporal attention block. Each block is formed by several stacked layers of the corresponding layer. They both employ self-attention mechanisms to obtain an efficient temporal graph representation from its neighboring and historical context information.

\subsubsection{Topological Attention Layer.}
The initial input for this layer is a set of brain network attributes $\left\{x_i\in\mathbb{R}^d,\forall i\in\mathcal{V}\right\}$ where $d$ is the dimension of time series. The output is a set of brain region representations $\left\{h_i\in\mathbb{R}^f,\forall i\in\mathcal{V}\right\}$ where $f$ is the dimension of captured topological properties.

Similar to graph attention networks(GAT)~\cite{ref_19},our topological attention layer is concerned with the near neighbors of the brain region $i$ by calculating attention weight from input brain region representations:

\begin{equation}
\begin{aligned}
\boldsymbol{h}_{i}&=\sigma\left(\sum_{u \in \mathcal{N}_{i}} \alpha_{i j} \boldsymbol{W} \boldsymbol{x}_{j}\right), \\ \alpha_{i j}&=\frac{\exp \left(\sigma\left(A_{i j} \cdot \left[\boldsymbol{W} \boldsymbol{x}_{j} \| \boldsymbol{W} \boldsymbol{x}_{i}\right]\right)\right)}{\sum_{w \in \mathcal{N}_{i}} \exp \left(\sigma\left(A_{w i} \cdot \left[\boldsymbol{W} \boldsymbol{x}_{w} \| \boldsymbol{W} \boldsymbol{x}_{i}\right]\right)\right)}
\end{aligned}
\end{equation}

Here $\mathcal{N}_i=\left\{j\in\mathcal{V}:(i,j)\in\mathcal{E}\right\}$ is the set of near neighbor of region $i$ which are linked by functional connection $A$; $W\in\mathbb{R}^{d\times f}$ is a weight transformation matrix for each region representations; $\sigma(\cdot)$ is  sigmoid activation function and $\|$ is the concatenation operation. The learnt coefficients $\alpha_{ij}$, which is computed by performing softmax on each neighbors, indicates the significance of brain region $i$ to region $j$.Note that  topological attention layer applies on brain region representation at a single timestamp, and multiple topological attention layer can calculate the entire time sequence in parallel.

\subsubsection{Temporal Attention Layer.}
Dynamically capturing constant changing patterns of brain networks is essential for dynamic community detection. When extracting the local timestamp features, it is critical to consider the influence of the global temporal context. The key question is how to capture the temporal alterations in brain networks structure throughout variety of time steps. Temporal attention layer is designed for tackling this issue with the help of the scaled dot-product attention~\cite{ref_20}. Its queries, keys,and values are being used to represent the attributes of input brain regions.

We define $H_s=\left\{h^1_s,h^2_s,...,h^T_s\right\}$ , a representation sequence of a brain region $s$ at continuous timestamps as input, where $T$ is the number of time steps. And the output of the layer is $Z_s=\left\{z^1_s,z^2_s,...,z^T_s\right\}$, a new brain network representation sequence for region $s$ at different timestamp.

Using $h^t_s$ as the query, temporal attention layer evaluate its historical representations, inquiring the temporal context of the neighborhood around region $s$. Hence, temporal self-attention allows the discovery of relationships between time-varying representations of a brain region across several time steps. Formally, the temporal attention layer is computed as:

\begin{equation}
\begin{aligned}
\boldsymbol{Z}_{s}&=\boldsymbol{\beta}_{\boldsymbol{s}}\left(\boldsymbol{H}_{s} \boldsymbol{W}_{s}\right),\\
\beta_{s}^{i j}=\frac{\exp \left(e_{s}^{i j}\right)}{\sum_{k=1}^{T} \exp \left(e_{s}^{i k}\right)}, \quad e_{s}^{i j}&=\left(\frac{\left(\left(\boldsymbol{H}_{s} \boldsymbol{W}_{q}\right)\left(\boldsymbol{H}_{s} \boldsymbol{W}_{k}\right)^{T}\right)_{i j}}{\sqrt{F^{\prime}}}\right)\\
\end{aligned}
\end{equation}

where $\beta_s\in\mathbb{R}^{T\times T}$ is the attention coefficient matrix computed by the query-key dot product attention operation;$W_q\in\mathbb{R}^{d\times f}$,$W_k\in\mathbb{R}^{d\times f}$ and $W_v\in\mathbb{R}^{d\times f}$ are linear projections matrices which transform representations into a particular space.

The two attention blocks are calculated in sequence to obtain the final temporal representation, $i,e.$, the output embeddings $Z$. And It is utilized to reconstruct the brain network topology in the decoder:

\begin{equation}
\hat{A}=\sigma(ZZ^T)
\end{equation}

$\hat{A}$ is the reconstructed brain functional connection and  $\sigma(\cdot)$ is  still sigmoid function.

The classic reconstruction loss is defined by the form of cross entropy:

\begin{equation}
\mathcal{L}_{RE}=\sum \mathbb{E}\left[A_{i j} \log \hat{A}_{i j}+\left(1-A_{i j}\right) \log \left(1-\hat{A}_{i j}\right)\right]
\end{equation}

\subsection{Adversarial Learning}
In this adversarial model, the main objective is enforcing brain network embeddings $Z$ to match the prior distribution. Other naive regularizers push the learned embeddings to conform to the Gaussian distribution rather than capture semantic diversity. As a result, conventional techniques to network embedding cannot effectively profit from adversarial learning. Therefore, we derive the previous distribution of communities by counting different kinds of modules in the functional brain network that have been confirmed by neuroscience. The adversarial model serves as a discriminator by using a three-layer fully connected network to identify whether a latent code drawn from the prior distribution  $p_{{z}'}$ (positive samples) or embeddings $z$ from the temporal graph encoder $E$ (negative samples). The regularizer will eventually enhance the embedding during the minimax competition between the encoder and the discriminator in the training phase.

The loss of the encoder(generator) $\mathcal{L}_G$ and discriminator $\mathcal{L}_D$ in the adversarial model, defined as follows:

\begin{equation}
\mathcal{L}_{G}=-\mathbb{E}_{x \sim p_{\text {data }}}\log \mathcal{D}_\phi \left(E_\psi\left(X,A\right)\right)
\end{equation}

\begin{equation}
\begin{aligned}
\mathcal{L}_{D}=&-\mathbb{E}_{z^{\prime} \sim p_{z^{\prime}}}\log \mathcal{D}_\phi({z}') \\
&-\mathbb{E}_{x \sim p_{\text {data }}}\log \left(1-\mathcal{D}_\phi \left(E_\psi\left(X,A\right)\right)\right)
\end{aligned}
\end{equation}

in this expression, ${z}'$ is a latent code sampled from the prior distribution $ p_{z^{\prime}}$ of empirically confirmed brain communities; $\mathcal{D}_\phi(\cdot)$ and $E_\psi(\cdot)$ is the above-mentioned discriminator and encoder.

Formally, the objective of this adversarial learning model can be indicated as a minmax criterion:

\begin{equation}
\begin{aligned}
\mathcal{L}_{\mathcal{AL}} &=\min _{G} \max _{D} \Theta\left(D, G\right) \\
&=\min _{G} \max _{D}\left( \mathbb{E}_{z^{\prime} \sim p_{z^{\prime}}}\log \mathcal{D}({z}';\phi)\right. \\
&+\left.\mathbb{E}_{x \sim p_{\text {data }}}\log \left(1-\mathcal{D} \left(E\left(X,A\right);\phi\right)\right)\right)
\end{aligned}
\end{equation}

\subsection{Measurable Modularity Loss}
Modularity maximization is a technique for community discovery that is commonly used in the detection of brain modules. A partition is regarded high quality (and so has a higher $Q$ score~\cite{ref_21}) conceptually if the communities it forms are more dense internally than would be predicted by chance. Thus, the partition that gets the maximum value of $Q$ is considered to be a good estimation of the community structure of a brain network. This intuition may be expressed as follows:

\begin{equation}
Q=\frac{1}{2 m} \sum_{i j}\left[A_{i j}-c_{i j}\right] \delta\left(\omega_{i}, \omega_{j}\right)
\end{equation}

here $a_{ij}$ indicates the number of functional connection between region $i$ and $j$; $c_{ij}=\frac{k_ik_j}{2m}$ denotes the estimated number of connections based on a null model where $k_i=\sum_j A_{ij}$ is a degree of the region $i$ and $2m=\sum_{ij}A_{ij}$ is overall amount of connections in the brain networks; $\delta(\omega_i,\omega_j)=1$ if  $\omega_i=\omega_j$, which means reigon $i$ and reigon $j$ are in the same community and 0 otherwise.

Inspired by~\cite{ref_22}, to develop a differentiable objective for optimizing the community assignment matrix $P=softmax(Z)\in\mathbb{R}^{N\times C}$ which represents a matrix of probabilities of brain region attribution to communities, the measurable modularity loss employed by our framework is defined as:

\begin{equation}
\mathcal{L}_{\mathrm{MM}}=\underbrace{-\frac{1}{\sum|A_{ij}|} \operatorname{tr}\left(\mathbf{P}^{\top} \mathbf{B P}\right)}_{\text {measurable modularity }}+\lambda\underbrace{
\left(\sum_{i}^{C}\left(\sum_{j}^{N} P_{i j}-\frac{1}{C}\right)^{2}\right)
}_{\text {regularization }}
\end{equation}

where the modularity matrix $B=A-\frac{dd^T}{2m}$; $C$ is the amount of communities and  $N$ is the number of regions in the brain networks. The regularization ensures that the model can identify communities of the predicted size.

Thus, the total loss for the encoder optimization in the train process to obtain better embeddings is sum of the above three loss terms, expressed as follows:

\begin{equation}
\mathcal{L}_{total}=\mathcal{L}_{RE}+\mathcal{L}_{G}+\mathcal{L}_{MM}
\end{equation}

\section{Experiments and Results}
In this part, we assess the performance of ATGRL in terms of both dynamic community detection and graph representation learning.

\subsection{Dataset Preparation and Implementation Details}

\subsubsection{Dataset Preparation.}
We obtained the dynamic brain network dataset required for the experiment by preprocessing long-term functional MRI images of experimental rats. The first preprocessing was carried out in MATLAB utilizing the Statistical Parametric Mapping 8 (SPM8) tool. To adjust for head motion, functional signals were aligned and unwrapped, and the mean motion-corrected image was coregistered with the high-resolution anatomical T2 image. Following that, the functional data were smoothed using a $3mm$ full-width at half-maximum (FWHM) isotropic Gaussian kernel. On the basis of the Wister rat brain atlas, 150 functional network areas were outlined. We used magnitude-squared coherence to assess the spectral relationship between regional time series, resulting in a $150\times 150$ functional connection matrix for each time step, whose members showed the intensity of functional connectivity between all pairs of areas.

\subsubsection{Implementation Details.}
ATGRL was implemented using pytorch backend. The training of the network was accelerated by one Nvidia GeForce RTX 2080 Ti. The training epoch was set at 500, while the learning rate was set to 0.001 during training. To minimize overfitting, Adam~\cite{Adam} was utilized as an optimizer with a weight decay of 0.01.  We trained the encoder with 2 topological attention layers and 2 temporal attention layers. We repeat all trials ten times and average the findings. For all datasets and approaches, we set the regularization value to 0.5 and the number of communities at 15.

\subsection{Dynamic Community Detection Performance }

\subsubsection{Baseline.}Our approach was compared against the following  two kinds of baselines:

\subsubsection{GAE.}~\cite{ref_23}is recently the most common autoencoder-based unsupervised framework for graph data, in which the encoder is composed of two-layer graph convolutional networks to leverage topological information.

\subsubsection{ARGA.}~\cite{ref_29}is an adversarially regularized autoencoder method that employs graph autoencoder to learn the representations, regularizes the latent codes, and forces the latent codes to match a prior distribution; differing from ours, it used simple Gaussian distribution as the prior distribution.

\subsubsection{Metrics.}
For graph-level metrics, we report average community conductance$\mathcal{C}$ and modularity$\mathcal{Q}$. For ground-truth label correlation analysis, we report normalized mutual information (NMI) between the community assignments and labels and pairwise F-1 score between all node pairs and their associated community pairs.

\begin{table}[]
\centering
\caption{Community detection performance on rat brain networks dataset by graph conductance $\mathcal{C}$, modularity $\mathcal{Q}$, NMI, and pairwise F1 measure.}\label{tab1}
\resizebox{\textwidth}{20mm}{\begin{tabular}{clp{1cm}p{1cm}p{1cm}p{1cm}}
\toprule
\multicolumn{1}{l}{}                   &                   & \multicolumn{4}{c}{\bfseries Metrics(\%)} \\ \cline{3-6}
\multicolumn{2}{c}{\bfseries method}  & $\mathcal{C}$        &$\mathcal{Q}$        & NMI      & F1      \\ \hline
\multirow{2}{*}{K-means based}             & GAE+K-means           & 72.4     & 13.6     & 50.7     & 30.6    \\
                                       & ARGA+K-means           & \bfseries75.0     & 22.3     & 49.6     & 46.3    \\ \hline
\multirow{3}{*}{Modularity loss based} & GAE+$\mathcal{L}_{MM}$              & 34.4     & 64.0     & 51.5     & 61.5    \\
                                       & Ours(GCN-encoder) & 21.5     & 59.7     & 45.7     & 58.2    \\ \cline{2-6}
                                       & Ours              & 36.0     & \bfseries68.3     & \bfseries68.9     & \bfseries63.3    \\
\bottomrule
\end{tabular}}
\end{table}

\subsubsection{Ablation Study.}
As indicated in Table.\ref{tab1}, we conducted ablation research on community detection to evaluate the effectiveness of our proposed encoder and adversarial learning, and three significant outcomes were achieved: 1) In the comparison of graph-level metrics, the k-means based method showed impressive performance on community conductance and the modularity loss based method did better than it on community modularity. This is due to the fact that the two algorithms are fundamentally different in terms of optimization; modularity loss originates with the goal of maximizing modularity. 2) The approach with adversarial regularizer is generally performed well; it represents that adversarial learning does play its role as an auxiliary to graph representation learning. 3) Our algorithm that replaced the proposed encoder with a two-layer graph convolution encoder performs worse; it shows in some way that our proposed encoder may learn better embeddings to make it perform well.

\begin{figure}
\includegraphics[width=\textwidth]{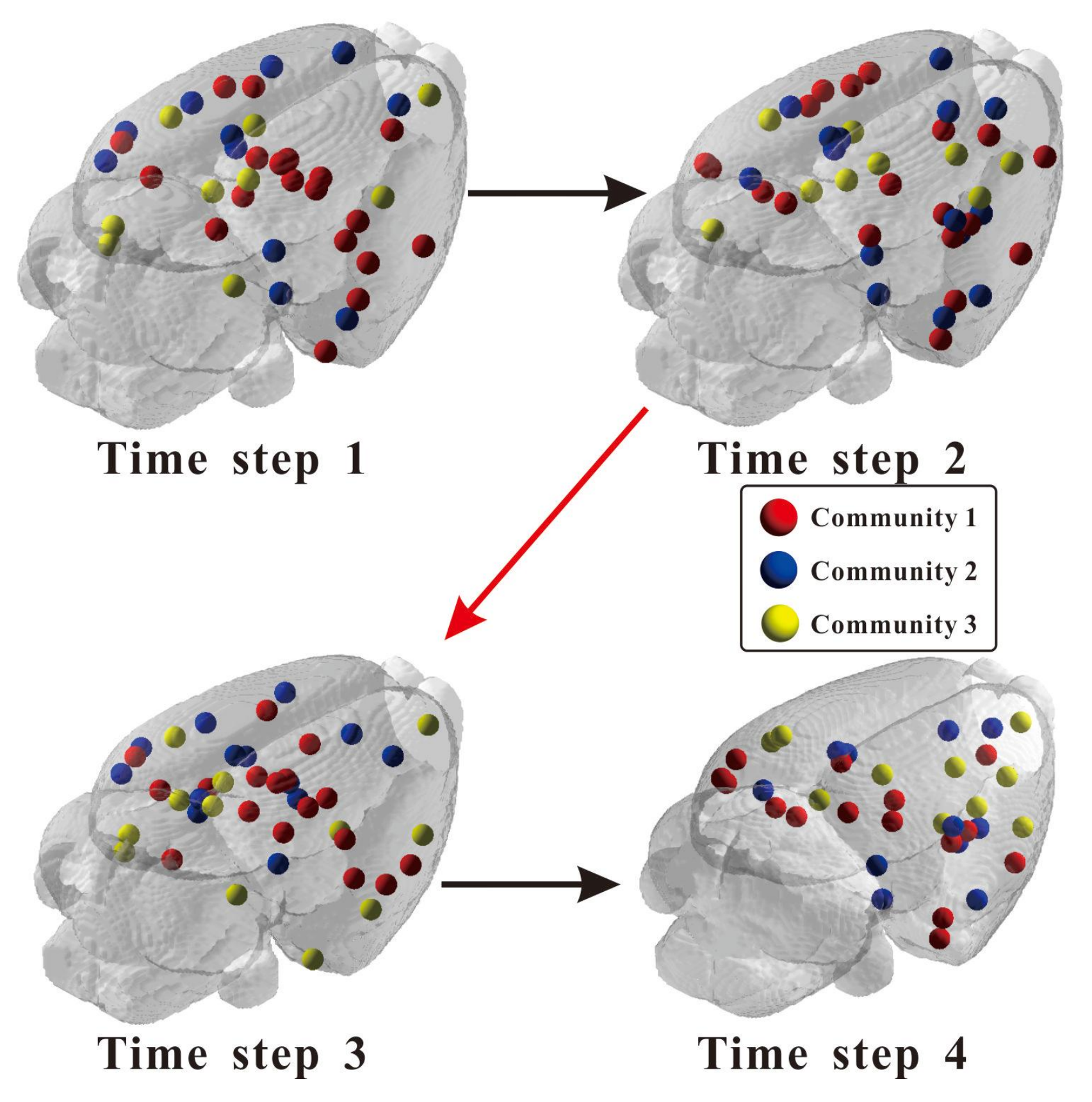}
\caption{Visualization of dynamic community detection performance within four time steps.} \label{fig2}
\end{figure}

\subsubsection{Visualization of Dynamic Community Detection.}
We illustrated our result of dynamic community detection by Fig.\ref{fig2}. It shows the changes in the positional distribution of the three major brain communities detected by our approach with increasing time steps. We can see that there is no significant change in the distribution of brain communities at time steps 1 to 2, but there is a more remarkable change at time steps 2 to 3. It is because the rats in the original dataset did change their brain network properties and topology due to experimental factors. Therefore, the outcomes of the experiment are in line with the neuroscientific truth.

\begin{figure}
\includegraphics[width=\textwidth]{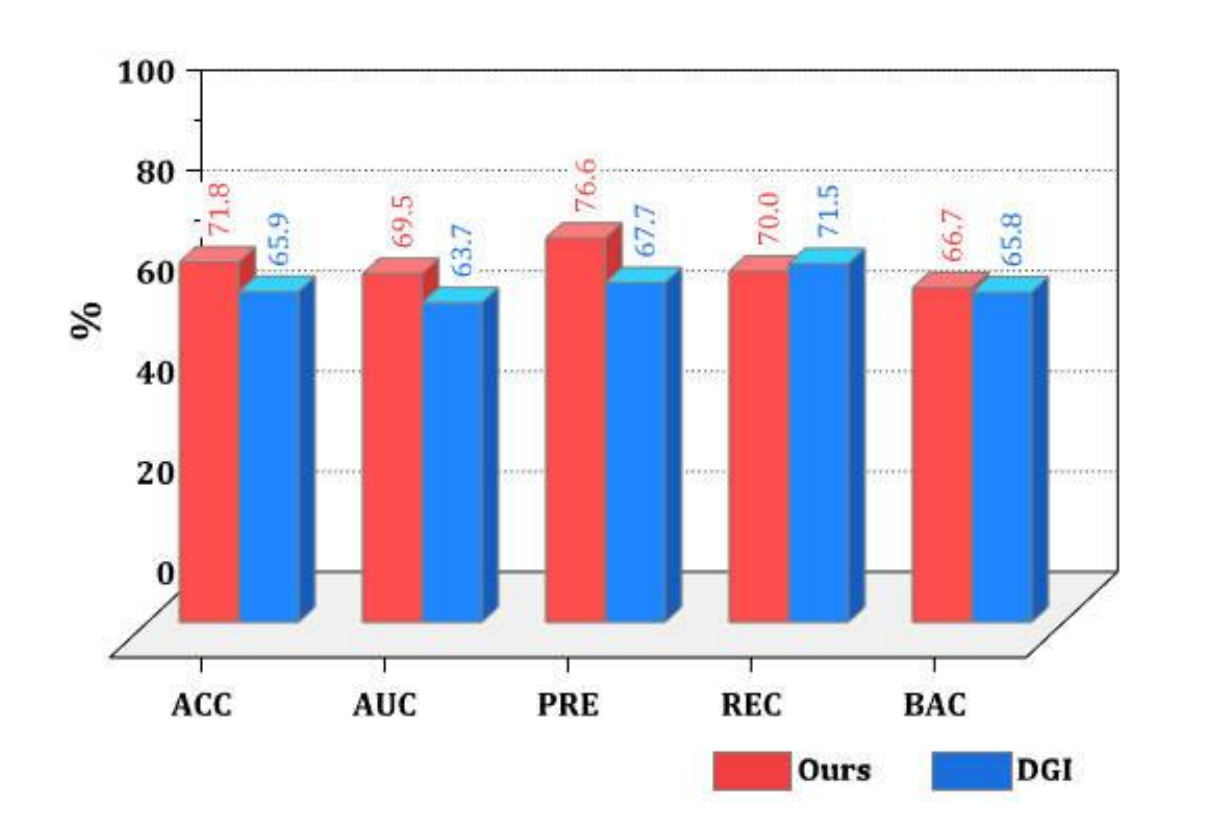}
\caption{Classification result of the proposed graph representaion learning and the competing method on baseline.} \label{fig3}
\end{figure}

\subsection{Graph Representation Learning Performance}

We grouped the rat data collected before and after the severe change into two groups and verified whether the model learned efficient graph representations by competing with the state-of-the-art graph representation learning model on classification performance.

\subsubsection{Competing Methods.}
DGI~\cite{ref_30} highlights the importance of cluster and representation learning in combination. We learn unsupervised graph representation with DGI and two algorithms both run SVM on the final representations as the classifier.

\subsubsection{Metrics.}
Evaluation of diagnostic performance is based on quantitative measures in five key areas: To summarize:  1) accuracy (ACC); 2) area under receiver operating characteristic curve (AUC); 3) Precision (PRE); 4) Recall (REC); and 5) balanced accuracy (BAC). Our suggested technique is being evaluated using leave-oneout cross-validation (LOOCV), since we only have a small quantity of data. One of the $N$ individuals is omitted from the testing process, and the $N-1$ subjects that remain are used for training purposes only. It is the greedy search that sets the hyperparameters in each technique to the optimum values.

\subsubsection{Prediction Results.}
As demonstrated in Fig.\ref{fig3}, our approach achieved generally better results on classification performance. In a respect, it verifies that the representations obtained by our method are more strong in the unsupervised learning process.

\section{Conclusion}
In this research, we propose a novel framework called Adversarial Temporal Graph Representation Learning (ATGRL) for introducing community detection into a deep graph representation learning process directed by an adversarial regularizer. In addition to using temporal graph attention encoder to merge input spatial topology features and temporal contextual representation to represent latent variables, adversarial training with a neuroscientific prior is used to deconstruct the embedding space in the ATGRL framework. Our method outperformed two unsupervised deep embedding and community identification approaches in dynamic brain network datasets. And we obtained better results than the comparison method when using the obtained graph representations for classification, indicating that there is an advantage in graph representation learning that may yield better graph embeddings in the latent space. Detailed model discussions were conducted to investigate the proposed ATGRL and the superiority of the encoder and the adversarial regularizer.

\end{document}